\documentclass[aps,prb,twocolumn,superscriptaddress,showpacs,floatfix]{revtex4}

\newcommand{\vk}{\mathbf{k}}
\newcommand{\vko}{\mathbf{k}_{0}}
\newcommand{\ko}{k_{0}}

\newcommand{\vv}{\mathbf{v}}
\newcommand{\vvr}{\mathbf{r}}

\newcommand{\vg}{v_{g}}
\newcommand{\vvg}{\vv_{g}}
\newcommand{\om}{\omega}
\newcommand{\al}{\alpha}
\newcommand{\be}{\beta}
\newcommand{\ga}{\gamma}
\newcommand{\sig}{\sigma}
\newcommand{\omo}{\om_{0}}
\newcommand{\Urt}{U(\vvr,t)}
\newcommand{\Uxt}{U(x,t)}
\newcommand{\pw}[1]{\mathrm{e}^{i#1}}
\newcommand{\pwm}[1]{\mathrm{e}^{-i#1}}

\newcommand{\eo}{\varepsilon}
\newcommand{\ek}{\epsilon_{k}}
\newcommand{\dn}{\varepsilon_{n}}
\newcommand{\dm}{\varepsilon_{m}}
\newcommand{\an}{a_{n}}
\newcommand{\ad}{a_{n}^{\dag}}
\newcommand{\bk}{b_{k}}
\newcommand{\bkd}{b_{k}^{\dag}}
\newcommand{\kmin}{k_\mathrm{min}}
\newcommand{\kmax}{k_\mathrm{max}}
\newcommand{\dk}{\delta k}
\newcommand{\WFp}{\Psi_{\mathrm{p}}}
\newcommand{\WFe}{\Psi_{\mathrm{e}}}
\newcommand{\WFps}{|\WFp|^{2}}
\newcommand{\WF}{\Psi}

\newcommand{\vWFt}{|\WF(t)\rangle}
\newcommand{\vWFo}{|\WF^{0}\rangle}
\newcommand{\vWFot}{|\WF^{0}(t)\rangle}
\newcommand{\vWFio}{|\WF^{0}_{i}\rangle}
\newcommand{\vWFi}{|\WF_{i}\rangle}
\newcommand{\eie}{\pwm{E_{i}t/\hbar}}
\newcommand{\eieo}{\pwm{E^{0}_{i}t/\hbar}}

\usepackage{graphicx}
\usepackage{amsmath}
\usepackage{latexsym}
\begin{document}

\title{On the nature and dynamics of low-energy cavity polaritons}

\author{V.~M.~Agranovich}
\affiliation{UTD-NanoTech Institute, The University of Texas at
Dallas, Richardson, TX 75083} \affiliation{Institute of
Spectroscopy, Russian Academy of Science, Troitsk, Moscow}
\author{Yu.~N.~Gartstein}
\affiliation{Department of Physics, The University of Texas at
Dallas, Richardson, TX 75083}

\date{\today}

\begin{abstract}
Low-energy polaritons in semiconductor microcavities are important
for many processes such as, e.g., polariton condensation. Organic
microcavities frequently feature both strong exciton-photon
coupling and substantial scattering in the exciton subsystem.
Low-energy polaritons possessing small or vanishing group
velocities are especially susceptible to the effects of such
scattering that can render them strongly localized. We compare the
time evolution of low-energy wave packets in perfect microcavities
and in a model 1$d$ cavity with diagonal disorder to illustrate
this localization of polaritons and to draw attention to the need
to explore its consequences for the kinetics and collective
properties of polaritons.
\end{abstract}

\pacs{78.66.Qn, 78.40.Me, 71.36.+c}

\maketitle

\section{Introduction}
Planar semiconductor microcavities have attracted much attention
as they provide a method to enhance and control the interaction
between the light and electronic excitations. When the microcavity
mode (cavity photon) is resonant with the excitonic transition,
two different regimes can be distinguished based on the
competition between the processes of the exciton-photon coupling
and damping (both photon damping and exciton dephasing). The weak
coupling regime corresponds to the damping prevailing over the
light-matter interaction, and the latter then simply modifies the
radiative decay rate and the emission angular pattern of the
cavity mode. In contrast, in the strong coupling regime the
damping processes are weak in comparison with the exciton-photon
interaction, and the true eigenstates of the system are mixed
exciton-photon states, cavity polaritons. This particularly
results in the appearance of the gap in the spectrum of the
excitations whose magnitude is established by Rabi (splitting)
energy. In inorganic semiconductors the strong coupling regime has
been investigated extensively, both experimentally and
theoretically,\cite{burweis95,weisrar96,skol98,koch99,Kavokin,Bass03,weisben06}
and the dynamics of microcavity polaritons is now reasonably well
understood.\cite{bottle} These studies continue to expand because
of the prospect of important applications  such as the polariton
laser related to the polariton condensation in the lowest energy
state.\cite{imam96,yama00,savv00,stev00,savv00a,alex01,deng02,Kavokin,rubo03,laus04}

In another development, organic materials have been utilized in
microcavities as optically active semiconductors. In many organic
materials excitons are known to be small-radius states, Frenkel
excitons, which interact much stronger with photons than
large-radius Wannier-Mott excitons in inorganic semiconductors.
The cavity polaritons, therefore, may exhibit much larger Rabi
splittings on the order of few hundreds of meV; large splittings
have in fact been observed
experimentally.\cite{lidz98,lidz99,tart01,hobs02,scho02,lidz02,taka03,lidz02a,lidz03,holm04,holm05}
At the same time, Frenkel excitons typically also feature
substantially stronger interactions with phonons and disorder -
electronic resonances in both disordered and crystalline organic
systems are frequently found rather broad and dispersionless. It
is thus likely that manifestations of exciton-polaritons in
organic microcavities could be quite different from the
corresponding inorganic counterparts.

In this paper we are concerned with the nature and dynamics of the
low energy exciton-polaritons in organic microcavities, the states
of particular importance for the problem of polariton
condensation. Our goal here is to illustrate some qualitative
features of the dynamics in both perfect and disordered systems
and, thereby, to draw attention to the need of more detailed
experimental and theoretical studies to elucidate conditions for
the polariton laser operation based on organic systems.

The bare planar cavity photons are coherent wave excitations with
a continuous spectrum and whose energy $\epsilon (\vk)=\ek$
depends on the magnitude $k=|\vk|$ of the 2$d$ wave vector $\vk$:
\begin{equation}\label{phot}
\ek=\left(\Delta^2 + \hbar^2 c^2 k^2 /\epsilon \right)^{1/2},
\end{equation}
where $\Delta$ is the cutoff energy for the lowest transverse
quantization photon branch we restrict our attention to, $c$ is
the speed of light and $\epsilon$ the appropriate dielectric
constant.

\begin{figure}
\includegraphics[scale=0.7]{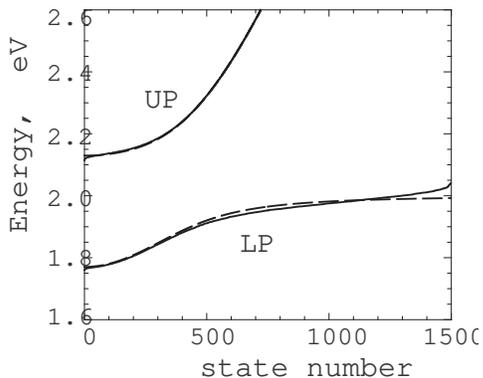}
\caption{\label{Spectrum}The energy spectrum of the polaritonic
eigenstates in a 1$d$ model microcavity  described by the
Hamiltonian (\ref{h1}) with $N=1500$ molecular sites and $N=1500$
photon modes. Parameters of the systems are as follows: average
exciton energy $\eo=2$ eV, cavity photon cutoff energy
$\Delta=1.9$ eV, dielectric constant $\epsilon=3$, the
exciton-photon interaction energy $\ga=0.15$ eV. The dashed lines
show the energy eigenvalues in the system without exciton
disorder, $\sig=0$, the solid lines correspond to the spectrum in
the system with disorder, $\sig=0.03$ eV. Here the energies are
shown as a function of state ``number'' sorted in an increasing
energy order, separately for the lower (LP) and upper (UP)
polariton branches. In the system without disorder, the state
numbers would be immediately convertible to the corresponding wave
vectors. On this scale, the UP branches of perfect and disordered
systems are hardly distinguishable.}
\end{figure}

In the vicinity of the excitonic resonance, $\ek \approx \eo$
($\eo$ being the exciton energy), strong exciton-photon coupling
leads to the formation of new mixed states of exciton-polaritons
whose two-branch ($E_{\pm}$) energy spectrum  features a gap as,
e.g., illustrated in Fig.~\ref{Spectrum}. Especially interesting
are systems with detuning $|\Delta-\eo|$ small in comparison with
the Rabi splitting. In the absence of the exciton scattering
processes, cavity polaritons are also clearly coherent excitations
that can be well characterized by wave vectors $\vk$ and have
energies $E_{\pm}(k)$. Frenkel exciton scattering (due to phonons
and/or disorder) in organic systems with weak intermolecular
interactions results in the exciton localization: excitons
propagate not as coherent wave packets but by hopping; in such a
regime the wave vector $\vk$ is no longer a ``good'' quantum
number.

As we are actually interested in lowest energy polariton states,
we will constrain further discussion mostly to the lower polariton
(LP) branch. One can quickly notice that, since the photon energy
(\ref{phot}) rapidly increases with $k$, higher-$k$ photon states
would be only very weakly interacting with exciton states.
Therefore a very large number of eigenstates of the system with
energies close the upper end of the LP branch (see
Fig.~\ref{Spectrum}) are essentially reflective of the bare
localized exciton states with an incoherent propagation mode. For
the remaining fewer and lower energy states of the LP branch,
however, the exciton-photon interaction is strong and one is then
faced with an interesting interplay of the bare localized nature
of the ``exciton part'' of the polariton and the bare coherent
character of its ``photon part''.

Transparent physical arguments were used in Ref.~\onlinecite{ALL}
involving the indeterminacy (broadening) of the \textit{polariton}
wave vector $k$ owing to the exciton dephasing from the general
relation:\cite{peierls}
\begin{equation}\label{dk}
\dk=\frac{dk}{dE}\,\delta E(k)=\frac{\delta E(k)}{\hbar \vg(k)},
\end{equation}
where $\vg (k)$ is the group velocity of parent polaritons in the
perfect systen and $\delta E(k)$ is the energy broadening due to
scattering. Based on Eq.~(\ref{dk}), one can at least distinguish
polaritonic states where $k$ is relatively well-defined in the
sense of
\begin{equation}\label{dk1}
\dk \ll k.
\end{equation}
If $\delta E(k)$ is weakly $k$-dependent, then, evidently from
Eq.~(\ref{dk}), condition (\ref{dk1}) would be necessarily
violated in regions of the spectrum where the group velocity
$\vg(k)$ vanishes. For the LP branch, as seen in
Fig.~\ref{Spectrum}, this occurs both at its lower- and
higher-energy ends. Reference \onlinecite{ALL} provided estimates
of the corresponding ``end-points'' $\kmin$ and $\kmax$, where
$\dk \simeq k$, for organic planar microcavities within the
macroscopic electrodynamics description of polaritons. It was then
anticipated that exciton scattering would render eigenstates
corresponding to parent polaritons with $k < \kmin$ and $k >
\kmax$ spatially localized in accordance with uncertainty
relations like $\delta x\dk_{x} \sim 1$. As we discussed above, at
the higher-energy end ($k > \kmax$) of the LP branch, the
eigenstates are practically bare exciton states in nature. At the
lower-energy end ($k < \kmin$), one would deal with localized
polaritons having comparable exciton and photon contents,
particularly for the detuning $|\Delta-\eo|$ small with respect to
the Rabi splitting . Further numerical calculations\cite{ML,AL}
for 1$d$ microcavities with diagonal exciton disorder confirmed
this qualitative picture for the polariton states. Below, we will
use a similar 1$d$ microcavity model to illustrate the nature of
the low-energy LP states as well as time evolution of low-energy
wave packets.

\section{Dynamics of low-energy wave packets in perfect microcavities}

Before proceeding with a model analysis of polaritons in a
disordered system, we will briefly discuss the time evolution of
wave packets in perfect microcavities where all polaritons are
coherent states well characterized by their wave vectors. Not only
will this establish a comparative benchmark but is useful in
itself as such dynamics reflects features of the polariton
spectrum, and hence of the exciton-photon hybridization.

Of course, specific features of the low-energy wave packets stem
from the fact that the polariton dispersion near the bottom of the
LP branch ($\vk \simeq 0$) is manifestly parabolic:
\begin{equation}\label{disp}
\om(\vk)\simeq \omo +\alpha k^{2}, \ \ \ \alpha=\hbar/2M,
\end{equation}
$M$ being the cavity polariton effective mass, which makes the
broadening of wave packets a relevant factor. Consider now a wave
packet formed with the states close to the branch bottom:
\begin{equation}\label{pack2d}
\Urt=\pwm{\omo t} \int d\vk \, A(\vk) \, \pw{\left(\vk\cdot\vvr -
\al k^2 t \right)}.
\end{equation}
It is convenient to choose the weight amplitude function $A(\vk)$
Gaussian: $A(\vk)=\left(\be/2\pi^3\right)^{1/2} \exp\left[-\be
(\vk-\vko)^{2}\right]$, centered at wave vector $\vko$. With this
amplitude function, Eq.~(\ref{pack2d}) yields
\begin{equation}\label{pack2da}
\left|\Urt\right|^2 = C(t)
 \,\exp\left[- \frac{\be (\vvr-2\al \vko t)^2}{2(\be^2 +
\al^2 t^2)} \right]
\end{equation}
for the time evolution of the spatial ``intensity'' of the wave
packet,
$$
C(t)=\frac{\be}{2\pi(\be^2+\al^2 t^2)}.
$$
Equation (\ref{pack2da}) describes a Gaussian-shaped wave packet
in 2$d$ whose center
$$
\vvr_{c}(t)=\vvg t, \ \ \ \vvg=2\al \vko,
$$
moves with the group velocity consistent with the dispersion
(\ref{disp}), and whose linear width increases with time in
accordance with the 1$d$ variance
\begin{equation}\label{var}
s(t)=\left(\be + \al^2 t^2/\be \right)^{1/2}.
\end{equation}
The total energy in the wave packet is conserved: with our choice
of the amplitude function,
$$
\int d\vvr \left|\Urt\right|^2 =1.
$$

Of course, Eq.~(\ref{pack2da}) can be derived as a product of two
independent 1$d$ normalized evolutions such as
\begin{equation}\label{pak1da}
\left|\Uxt\right|^2 =\sqrt{C(t)} \,\exp\left[- \frac{\be (x-2\al
\ko t)^2}{2(\be^2 + \al^2 t^2)} \right],
\end{equation}
which we will be relevant in our discussion of 1$d$ microcavities,
in this case the 1$d$ packet amplitude function
\begin{equation}\label{ampl1}
A(k)=\left(\be/2\pi^3\right)^{1/4} \exp\left[-\be
(k-\ko)^{2}\right].
\end{equation}

The spatial broadening (\ref{var}) features an initial value of
$\be^{1/2}$ and a characteristic time $t_{b}=\be/\al$ such that,
at times $t \gg t_{b}$, the variance grows linearly with velocity
$v_{b}=\al \be^{-1/2}$. To appreciate the scales, some rough
estimates can be made. So for the effective polariton mass $M$ on
the order of $10^{-5}\, m_{0}$ ($m_{0}$ being the vacuum electron
mass), parameter $\al=\hbar/2M \simeq 5\cdot 10^{4}$ cm$^2$/s.
Estimates in Ref.~\onlinecite{ALL} made with the Rabi splitting
and detuning $\sim 100$ meV yielded for microcavities with
disordered organics $\kmin \sim 10^{4}$ cm$^{-1}$. Then for the
wave packets satisfying $1 \lesssim \be^{1/2}\kmin \lesssim 10$,
the characteristic time would be $0.2 \lesssim t_{b} \lesssim 20$
ps and the corresponding velocity $5\cdot 10^8 \gtrsim v_{b}
\gtrsim 5\cdot 10^7$ cm/s. In our 1$d$ numerical example below we
will use the value of parameter $\be$ within the segment just
discussed.

We note that by changing physical parameters of the microcavity
and organic material, as well as conditions for the polariton
excitation, one can influence the dynamics described above. One
should also be aware that the evolution times are limited by the
actual life times $\tau$ of small wave-vector cavity polaritons.
Long life times $\tau$ on the order of 10 ps can be achieved only
in microcavities with high quality factors $Q=\om \tau$.

\section{Time evolution in a 1$d$ microcavity with diagonal disorder}

Finding polariton states in disordered planar microcavities
microscopically is a difficult task which we are not attempting in
this paper. As a first excursion into the study of disorder
effects on polariton dynamics, here we will follow
Ref.~\onlinecite{ML} to explore the dynamics in a simpler
microscopic model of a 1$d$ microcavity. Such microcavities are
interesting in themselves and can have experimental realizations;
from the results known in the theory of disordered
systems,\cite{lee1985} one can also anticipate that certain
qualitative features may be common for 1$d$ and 2$d$ systems.

The microscopic model we study is set up in the following
Hamiltonian:
\begin{eqnarray}
H & = & \sum_{n} (\eo + \dn)\ad\an + \sum_{k} \ek \bkd\bk
\nonumber \\
& + & \ga \sum_{nk} \sqrt{\frac{\eo}{N\ek}} \left(\pw{kna}\,\ad\bk
+ \pwm{kna}\,\bkd\an \right) \label{h1}.
\end{eqnarray}
It consists of a lattice of $N$ ``molecular sites'' spaced by
distance $a$ and comprises the exciton part ($\an$ is the exciton
annihilation operator on the site $n$), photon part ($\bk$ is the
photon annihilation operator with the wave vector $k$ and a given
polarization) as well as the ordinary exciton-photon interaction.
The cavity photon energy $\ek$ is defined by Eq.~(\ref{phot}),
$\eo$ represents the average exciton energy, while $\dn$ the
on-site exciton energy fluctuations. We will use uncorrelated
normally distributed $\dn$ with the zero mean and the variance
$\sig$:
\begin{equation}\label{dis}
\langle \dn\dm \rangle = \sig^2 \delta_{nm}.
\end{equation}
The exciton-photon interaction is written in such a form that
$2\ga$ yields the Rabi splitting energy in the perfect system. We
chose to use the same number $N$ of photon modes, the wave vectors
$k$ are discrete with $2\pi/Na$ increments. Our approach is to
straightforwardly find the normalized polariton eigenstates
$\vWFi$ ($i$ is the state index) of the Hamiltonian (\ref{h1}) and
then use them in the site-coordinate representation:
\begin{equation}\label{wf}
\WF (n) = \left(\WFp (n), \, \WFe (n) \right),  \ \ \sum_{n}|\WF
(n)|^2 = 1,
\end{equation}
where $\WFp$ and $\WFe$ respectively describe the photon and
exciton parts of the polariton wave function and $n$ denotes the
$n$th site.

We have tried various numerical parameters in the model
Hamiltonian with the results being qualitatively consistent; the
parameters we exploit in this paper have been chosen, on one hand,
to be reasonably comparable with the experimental data in the
output and, on the other hand, to better illustrate our point
within a practical computational effort. It should be kept in mind
though that we consider a model system and the numerical values of
results may differ, likely within an order of magnitude, for
various systems.

The numerical parameters are indicated in the caption to
Fig.~\ref{Spectrum} and have been used to calculate the
eigenstates of the Hamiltonian (\ref{h1}) with $N=1500$ for a
cavity of the physical length $L=Na=150$ $\mu$m and a small
negative detuning $(\Delta - \eo)=-0.1$ eV. Figure \ref{Spectrum}
compares the energy spectra in the perfect microcavity and in the
cavity with one realization of the excitonic disorder,
Eq.~(\ref{dis}), $\sig=0.03$ eV. It is apparent that the effect of
this amount of disorder on the polariton energy spectrum
\textit{per se} is relatively small, except in the higher-energy
region of the LP branch where eigenstates, as we discussed, are
practically of a pure exciton nature.

\begin{figure}
\includegraphics[scale=0.7]{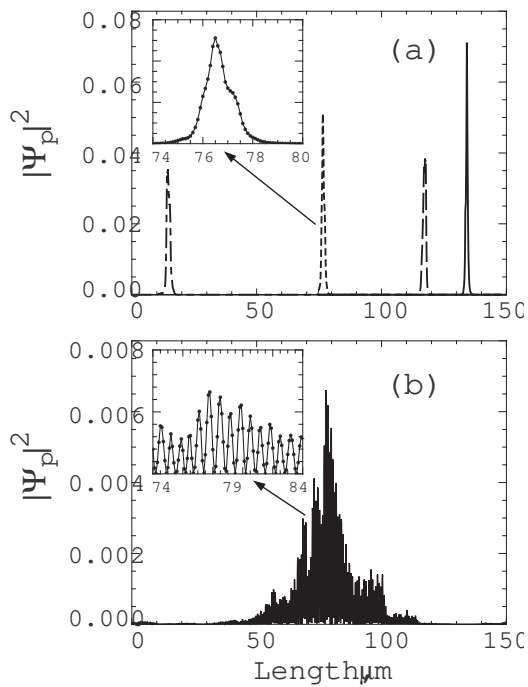}
\caption{\label{Wfunctions}Examples of the spatial structure of
the the photon part $\WFps$ of the polariton eigenstates in a 1$d$
microcavity with disorder. (a) Four states, shown by different
lines, from the very bottom of the LP polariton branch with
energies within the range of 1.76-1.77 eV (see the spectrum in
Fig.~\ref{Spectrum}). The inset shows one of these states in more
detail. Dots in the inset correspond to the sites of the
underlying lattice. (b) One state with a higher energy close to
1.82 eV. The inset shows part of the spatial structure of this
state in more detail. Dots correspond to the lattice sites and
spatial oscillations of the wave function are clearly seen.}
\end{figure}

\begin{figure}
\includegraphics[scale=0.7]{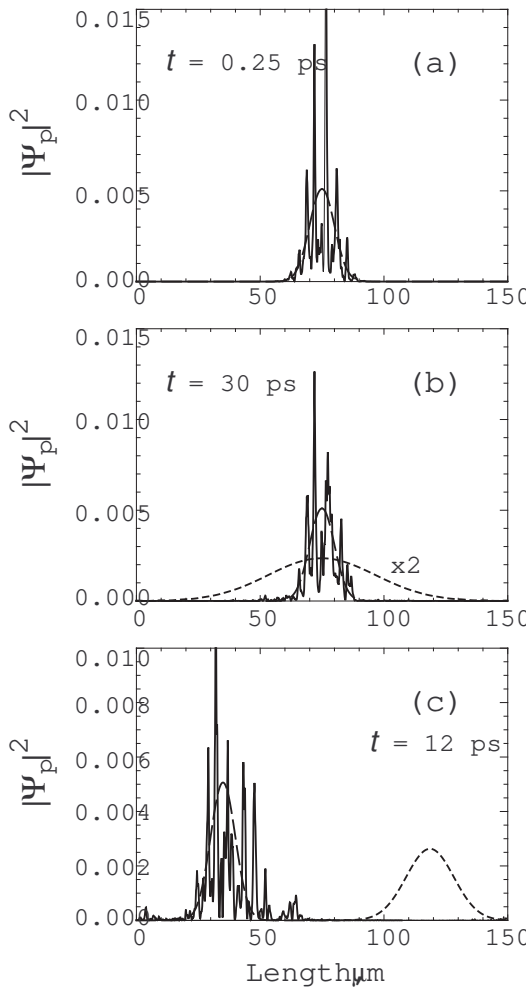}
\caption{\label{Evol}Examples of the the time evolution of
spatially identical wave packets built out of the polariton
eigenstates of a perfect 1$d$ microcavity as in Eq.~(\ref{ampl1})
with the parameter $\be^{1/2}=5$ $\mu$m. For panels (a) and (b),
the initial packet has zero total momentum, $\ko=0$; for panel (c)
the initial packet has a finite momentum determined by $\ko=10^4$
cm$^{-1}$. Only the photon part $\WFps$ of the polariton wave
function is displayed. The initial packets are shown by
long-dashed lines, results of the evolution after indicated times
$t$ are shown by solid lines for the disordered system and by
short-dashed lines for the perfect microcavity (except in panel
(a), where the latter practically coincides with the initial
packet.)}
\end{figure}

The lower-energy part of the LP branch, however, corresponds to
the polariton states $\WF$ (\ref{wf}) in which the exciton and
photon are strongly coupled ($\ga=0.15 > |\Delta -\eo|=0.1$ eV)
with comparable weight contributions in $\WFp$ and $\WFe$. A
dramatic effect of the disorder is in the strongly localized
character of the polaritonic eigenstates near the bottom of the LP
branch, as illustrated in Fig.~\ref{Wfunctions}(a) (needless to
say that the same behavior is observed for the states near the
bottom of the UP branch which we we are not concerned with in this
paper). Of course, both the photon $\WFp$ and exciton $\WFe$ parts
of a localized polariton state are localized on the same spatial
scale, however the exciton part of the spatial wave function is
more ``wiggly'' reflecting more the individual site energy
fluctuations. For better clarity, in both Figs.~\ref{Wfunctions}
and \ref{Evol} we show only the smoother behaving photon parts
$\WFp$. Panel (a) of Fig.~\ref{Wfunctions} displays examples of
$\WFps$ for four states from the very bottom of the LP branch in a
realization of the disordered system that are localized at
different locations of the 150 $\mu$m sample. The inset to this
panel shows the spatial structure of one of these states in more
detail; it demonstrates both the spatial scale $l$ of localization
in this energy range ($l \sim 1$ $\mu$m with the used parameters)
as well as a ``macroscopic'' size of the localized state in
comparison with the lattice spacing: $l \gg a=100$ nm.

The states at the bottom of the LP branch can be characterized as
strongly localized in the sense of $k l \lesssim 1$ where $k$ is a
typical wave vector of the parent polariton states in the perfect
system. This feature may be contrasted to the behavior at somewhat
higher energies and at higher $k
> \kmin$ of parent states, where the disorder-induced indeterminacy of
the $k$-vector becomes small satisfying Eq.~(\ref{dk1}) so that
$k$ would appear as a good quantum number. As is
known,\cite{lee1985} however, the multiple scattering should still
lead to spatial localization of the eigenstates, now on the
spatial scale $l$ such that $k l \gg 1$. Panel (b) of
Fig.~\ref{Wfunctions} illustrates the spatial structure of such a
state with much larger $l$ than in panel (a). The inset to panel
(b) shows that the wave function in this case, within the
localization length, exhibits multiple oscillations with a period
on the order of $1/k$, which produces the black appearance on the
scale of whole panel (b).

Having all the eigenstates of the system calculated, we are now in
a position to study the time evolution of an initial polariton
excitation, which we choose in the form of a wave packet $\vWFo$
built out of the low-energy polariton states $\vWFio$ of the
perfect system:
\begin{equation}\label{ini}
\vWFo = \sum_{i} A_{i}\,\vWFio =\sum_{i} B_{i}\,\vWFi.
\end{equation}
Polaritons  in the perfect system are ordinary plane waves and we
used a discretized analog of Eq.~(\ref{ampl1}) for the amplitude
function $A_{i}$, the result is a Gaussian-shaped wave packet as
illustrated in Fig.~\ref{Evol} by the long-dashed lines for the
photon part of the polariton wave function. Amplitudes $B_{i}$ in
Eq.~(\ref{ini}) are, on the other hand, expansion coefficients of
the same initial excitation over the eigenstates $\vWFi$ of the
system with disorder. The time evolution of the initial excitation
in the perfect system is then given by
\begin{equation}\label{evol0}
\vWFot = \sum_{i} A_{i}\,\eieo \, \vWFio,
\end{equation}
while the evolution in the disordered system by
\begin{equation}\label{evol1}
\vWFt = \sum_{i} B_{i}\,\eie \, \vWFi,
\end{equation}
where $E_{i}^{0}$ and $E_{i}$ are the respective eigenstate
energies.

Of course, the evolution of the low-energy wave packet
(\ref{evol0}) in the perfect microcavity takes place in accordance
with our continuum generic description in Eq.~(\ref{pak1da})
(barring small differences that may be caused by deviations from
the purely parabolic spectrum). This is clearly seen in panels (b)
and (c) of Fig.~(\ref{Evol}) where the photon part of $\vWFot$ at
indicated times $t$ is displayed by the short-dash lines: mere
broadening of the wave packet with no momentum ($\ko=0$) in panel
(b), and both broadening and translational displacement ($\ko \neq
0$) in panel (c). On the time scale of panel (a), the $\vWFot$
state has not practically evolved yet from $\vWFo$ and is not
shown on that panel.

The time evolution of exactly the same initial polariton packets
is drastically different in the disordered system, the
corresponding spatial patterns of the photon part of $\vWFt$ are
shown in Fig.~\ref{Evol} with solid lines. First of all, the
initial packet is quickly (faster than a fraction of ps)
transformed into a lumpy structure reflecting the multitude of the
localized polariton states within the spatial region of the
initial excitation. Note that in our illustration here we
intentionally chose the initial amplitude function (\ref{ampl1})
with the parameter $\be^{1/2}=5$ $\mu$m large enough for the
spatial size of the initial excitation to be much larger than the
size of the individual localized polaritons at these energies
(compare to Fig.~\ref{Wfunctions}(a)). Importantly, however, that,
while displaying some internal dynamics (likely resulting from the
overlap of various localized states), this lumpy structure does
not propagate well beyond the initial excitation region over
longer times. This is especially evident in comparison, when the
broadening and motion of the packets in the perfect system is
apparent (panels (b) and (c) of Fig.~\ref{Evol}). We have run
simulations over extended periods of time ($\sim 100$ ps) with the
result that $\vWFt$ remains essentially localized within the same
spatial region. Of course, some details of $\vWFt$ depend on the
initial excitation - see, e.g., a somewhat broader localization
region in Fig.~\ref{Evol}(c) for the initial excitation with an
initial momentum corresponding to $\ko=10^{4}$ cm$^{-1}$ but the
long-term localization in the disordered system appears robust in
all our simulations. It would be interesting to extend the
dynamical studies with participation of higher energy states such
as in Fig.~\ref{Wfunctions}(b), this is, however, beyond the scope
of the present paper.

\section{Concluding remarks}

The nature and dynamics of low-energy cavity polariton states are
important for various physical processes in microcavities,
particularly for the problem of condensation of polaritons into
the lowest energy state(s). As was demonstrated in
Ref.~\onlinecite{ALL}, low-energy polaritons in organic
microcavities should be especially susceptible to effects of
scattering/disorder in the exciton subsystem. The problem of
disorder effects on polaritons in organic microcavities appears
quite interesting as organic materials would typically feature
both strong exciton-photon coupling and substantial static and/or
dynamic exciton scattering. In this paper we have continued a line
of study in Ref.~\onlinecite{ML} to look in some more detail at
disorder effects on polaritons in a 1$d$ model microcavity. Our
numerical analysis has brought further evidence that low-energy
polariton states in organic microcavities can be strongly
localized in the sense of $l \lesssim \lambda$, where $l$ is the
spatial size of localized states and $\lambda$ the wave length of
parent polariton waves. (We have also found indications of weaker
localization at higher polariton energies in the sense of $l \gg
\lambda$.) Our illustrations have included demonstrations of
localization not only via the spatial appearance of polariton
eigenstates but also via the time evolution of different
low-energy wave packets.

On the physical grounds,\cite{ALL} one should expect that
low-energy polaritons in 2$d$ organic microcavities would also be
rendered strongly localized by disorder, as it would also follow
from the general ideas of the theory of
localization.\cite{lee1985} Further work on microscopic models of
2$d$ polariton systems is required to quantify their localization
regimes.

The strongly localized nature of low-energy polariton states
should affect many processes such as light scattering and
nonlinear phenomena as well as temperature-induced diffusion of
polaritons. Manifestations of the localized polariton statistics
(Frenkel excitons are paulions exhibiting properties intermediate
between fermi and bose particles) in the problem of condensation
also appear interesting and important.

We note that one can exercise an experimental control over the
degree of exciton-photon hybridization and disorder by modifying
the size of microcavity for various organic materials making such
systems a fertile ground for detailed experimental and theoretical
research into their physics.

And the last remark. While we have specifically discussed
exciton-photon polaritons in organic systems, it is clear that
some aspects have a generic character and could be applicable to
other systems. This, for instance, concerns inorganic
semiconductor microcavities. Both exciton-photon coupling and
magnitudes of disorder there, however, are much smaller, which
would likely make any localization effects relevant only at very
low temperatures. As one of recent examples of a very different
kind of systems, we will mention hybrid modes in chains of
noncontacting noble metal nanoparticles where the interaction of
photons and nanoparticles lead to
plasmon-polaritons.\cite{citrin2006}

\section{Acknowledgements}

VMA's work was supported by Russian Foundation of Basic Research
and Ministry of Science and Technology of Russian Federation. He
is also grateful to G.~C.~La Rocca for discussions. The authors
thank D.~Basko for reading and commenting on the manuscript.

\bibliography{cavpol}

\end{document}